\documentclass[twocolumn,showpacs,prb,superscriptaddress]{revtex4}
\usepackage{graphicx}
\usepackage{dcolumn}
\usepackage{amsmath}
\usepackage{amssymb}

\makeatletter
\def\btt#1{\texttt{\@backslashchar#1}}
\DeclareRobustCommand\bblash{\btt{\@backslashchar}} \makeatother

\begin{document}

\title{Josephson Effect in Noncentrosymmetric Superconductor Junctions}

\author{Yasuhiro Asano}
\affiliation{Department of Applied Physics and Center for Topological Science \& Technology,
Hokkaido University, Sapporo 060-8628, Japan}

\author{Satoshi Yamano}
\affiliation{Department of Applied Physics,
Hokkaido University, Sapporo 060-8628, Japan}

\date{\today}

\begin{abstract}
We discuss the Josephson current between two 
noncentrosymmetric superconductors. 
The coexistence of superconducting order parameters between 
spin-singlet $\Delta_{\text{S}}$ 
and helical $p$-wave spin-triplet $\Delta_{\text{T}}$
 enriches a variety of 
 low-temperature behavior of Josephson current depending on 
 their relative amplitudes.
We will show that characteristic behaviors of the Josephson current 
for $\Delta_{\text{S}} > \Delta_{\text{T}}$ are clearly different 
from those for $\Delta_{\text{S}} < \Delta_{\text{T}}$.
The topologically protected zero-energy surface bound states are responsible for 
the clear difference. We conclude that the Josephson current 
well reflects character of the topological surface states 
and the pairing symmetry of noncentrosymmetric superconductors.
\end{abstract}


\maketitle

\section{introduction}
Coexistence between the spin-singlet superconducting order parameters and spin-triplet one is 
the essential feature of noncentrosymmetric superconductors 
(NCS)\cite{bauer,togano,nishiyama,frigeri}. 
The absence of spatial inversion symmetry leads to spin-orbit coupling 
large enough to mix the spin-singlet component and spin-triplet one. 
The amplitude of the spin-singlet component $\Delta_{\text{S}}$ and that of 
the spin-triplet one $\Delta_{\text{T}}$
is a material parameter determined by the amplitude of spin-orbit coupling.
The Rashba type spin-orbit coupling induces the 
helical $p$-wave spin-triplet order parameter which is the topologically nontrivial 
superconducting state~\cite{frigeri,sato}.
There have been several studies on superconducting properties 
coexisting of the spin-singlet $s$-wave and the spin-triplet helical 
$p$-wave symmetries~\cite{tanaka09,vorontsov,fujimoto,yanase,iniotakis,linder,lu}. 
It is known that topologically protected states with linear dispersion appear 
at a surface of a NCS for $ \Delta_{\text{T}} > \Delta_{\text{S}}$. 
Recent papers, however, have suggested a mixed order parameter spin-singlet $d$-wave 
and spin-triplet $p$-wave symmetries~\cite{tanaka10,yada} which has been 
proposed for the interfacial superconductivity~\cite{reyren}.
Such pairing symmetry results in dispersionless surface bound state 
at the fermi level. 
A similar flat zero-energy surface states has also discussed 
in a NCS very recently~\cite{brydon}.
 Physical values originated from the bulk region of a superconductor 
such as the specific heat and spin susceptibility~\cite{frigeri} are 
expected to be interpolated from those in the two limits:
the pure spin-singlet case and the pure spin-triplet one. 
An open question is how physical values governed by the surface bound states behave 
as a function of 
the relative amplitude between $\Delta_{\text{T}}$ and $\Delta_{\text{S}}$. 
The present paper addresses this issue. 

The surface bound states of unconventional superconductors have been 
theoretically discussed
in heavy fermionic superconductors~\cite{buchholtz}, the polar state of 
$^3$He~\cite{hara}, and high-$T_c$ cuprates~\cite{hu,tanaka95}.
Experimentally, the presence of the surface bound states have been observed 
as the zero-bias anomaly~\cite{tanaka95,ya04} of the scanning tunneling spectroscopy (STS) of 
hole-doped~\cite{kashiwaya,wei} and electron-doped~\cite{biswas} high-$T_c$ cuprates. 
The zero-bias anomaly has been observed also in the differential conductance of 
ramp-edge junctions of hole-doped high-$T_c$ 
cuprates~\cite{iguchi} and grain boundary junction of electron-doped 
high-$T_c$ cuprates~\cite{chesca}. 
The presence of the surface bound states has been reinterpreted 
since the proposal for new classification of matter~\cite{kane}. 
The surface bound states at the zero energy are necessary to 
naturally connect a nontrivial topological integer number inside of 
an unconventional superconductor with the trivial topological 
number outside of the superconductor. 
The dispersion of the subgap 
states depends on the type of the topological number defined in 
superconductors. The chiral or helical superconductors 
give rise to dispersive surface bound states~\cite{qi}.
On the other hand, dispersionless zero-energy states 
are formed under $d_{x^2-y^2}$- and $p_{x}$-wave symmetries. 

 In direct current Josephson effect,
the surface bound states result in large $J_cR_N/(\Delta_0/e)$ values 
and the deviation of current-phase relationship from the sinusoidal function 
at low temperature~\cite{tanaka96,barash}, 
where $J_c$ is the critical Josephson current, $R_N$ is the normal resistance of 
a junction, and $\Delta_0$ is the amplitude of pair potential at the zero temperature.  
Such behavior is called low-temperature anomaly of Josephson current
and is known to be sensitive to spectra of surface bound states~\cite{tanaka96,barash,barash2,ya02,yakovenco}. So far the Josephson effect between
$s$-wave superconductor and a NCS has been reported~\cite{hayashi}. The low-temperature
anomaly of the Josephson current has never been discussed yet.

In this paper, we theoretically calculate the Josephson current between two 
NCS's~\cite{kaur,sumiyama}
based on a current formula~\cite{ya01} in terms of 
the Andreev reflection coefficients of junctions.
We assume a order parameter which is a mixture of the 
spin-singlet $s$-wave and the spin-triplet helical 
$p$-wave symmetries. 
We call such states as $s+p$ mixture. 
For $\Delta_{\text{T}} < \Delta_{\text{S}}$, 
the Josephson current saturates at low temperature as 
is described by the Ambegaokar-Baratoff formula~\cite{ambegaokar}. 
On the other hand for $\Delta_{\text{T}} > \Delta_{\text{S}}$,
the Josephson current increases logarithmically with 
decreasing temperature ($T$). 
The interfacial bound state causes the low-temperature anomaly. 
The characteristic 
behavior of the Josephson current does not changes gradually as a function of 
the relative amplitude between $\Delta_{\text{T}}$ and $\Delta_{\text{S}}$. 
The critical point $\Delta_{\text{T}} = \Delta_{\text{S}}$ clearly divides 
the qualitative feature of Josephson current. 

In addition to $s+p$ mixture, we also consider two types of mixed order 
parameter between the 
spin-singlet $d_{xy}$-wave and the spin-triplet helical $p$-wave 
symmetries. We call such state as $d+p$ mixture. 
The feature of Josephson current are well characterized by the relative amplitude between $\Delta_{\text{T}}$ and $\Delta_{\text{S}}$. 
In some cases, the Josephson current 
 increases as $1/T$ with decreasing temperature
due to dispersionless zero-energy state. 
We will discuss the physics behind such clear qualitative change of the Josephson 
effect in terms of topologically protected zero-energy surface states.
It is known that excitation of such surface bound states on superconductor 
are characterized by the Majorana fermion~\cite{wilczek,qi,fu}. 
Unusual phenomena peculiar to the Majorana fermion has been 
suggested theoretically~\cite{fu2,ya10,feigelman}.

This paper is organized as follows. In Sec.~II, we discuss a theoretical model
of Josephson junction consisting two NCS's. In Sec.~III, we show the calculated 
results of Josephson current for $s+p$ and $d+p$ mixtures. 
 We summarize this paper in Sec.~IV.
\begin{figure}[th]
\begin{center}
\includegraphics[width=8cm]{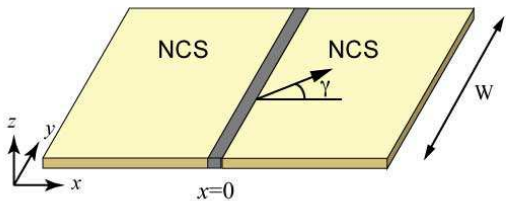}
\end{center}
\caption{(color online). A schematic picture of the Josephson junction. 
}
\label{fig1}
\end{figure}
\section{model}
Let us consider a Josephson junction between two NCS's as shown in Fig.~\ref{fig1}, 
where the electric current flows in the $x$ direction and the junction width in the 
$y$ direction is $W$. We apply the periodic boundary condition in the $y$ direction.
The Bogoliubov-de Gennes (BdG) Hamiltonian in momentum space reads
\begin{align}
H_{\textrm{BdG}}(\boldsymbol{k})=&\left[ \begin{array}{cc}
\hat{h}(\boldsymbol{k}) & \hat{\Delta}(\boldsymbol{k}) \\
-\hat{\Delta}^\ast(-\boldsymbol{k}) &-\hat{h}^\ast(-\boldsymbol{k})
\end{array}\right], \label{bdg}\\
\hat{h}(\boldsymbol{k})=& \xi_{\boldsymbol{k}} \hat{\sigma}_0
+\lambda\boldsymbol{g}\cdot\hat{\boldsymbol{\sigma} },\\
\xi_{\boldsymbol{k}}=&\frac{\hbar^2\boldsymbol{k}^2}{2m} -\mu,
\end{align}
where $\hat{\sigma}_j$ for $j=1-3$ are the Pauli matrices, $\hat{\sigma}_0$
is the unit matrix in spin space, $k_{x(y)}$ is the wavenumber in the $x(y)$ 
direction, $k_F$ is the Fermi wave number, $\mu$ is the chemical potential,
 and $\lambda$ is amplitude of the spin-orbit interaction. 
In this paper,  we assume that $\lambda \ll \mu$.
We consider the Rashba type spin-orbit coupling reflecting the noncentrosymmetry
along the $z$ direction (i.e., $\boldsymbol{g}=(k_y,-k_x,0)/k_F$). 
Correspondingly, we choose the $\boldsymbol{d}$-vector in the pair potential as
 $\boldsymbol{d} =\boldsymbol{g}$ as discussed in Ref.~\onlinecite{frigeri}. 
As a consequence, the spin-triplet part of the pair potential has 
the helical $p$-wave symmetry. 
In this paper, we consider three types of mixed order parameter as follows
\begin{align}
\hat{\Delta}(\boldsymbol{k})= \left\{ 
\begin{array}{ll}
 i\left( \Delta_{\text{T}}\boldsymbol{d}\cdot\boldsymbol{\sigma}
+\Delta_{\text{S}} \right) \hat{\sigma}_2 & {s+p }, \\
 i\left( \Delta_{\text{T}}\boldsymbol{d}\cdot\boldsymbol{\sigma}
+\Delta_{\text{S}}\sin 2\gamma \right)\hat{\sigma}_2 & {d+p} \; \text{ I}, \\
i\left( \Delta_{\text{T}}\boldsymbol{d}\cdot\boldsymbol{\sigma}
+\Delta_{\text{S}} \right)\sin 2\gamma \hat{\sigma}_2  & d+p \text{ II}, 
\end{array}\right.
\end{align}
where $\gamma$ is the incident angle of a quasiparticle as shown in Fig.~\ref{fig1}
and $e^{i\gamma}=(k_x+ik_y)/k_F$.
The first one consists of $s$-wave singlet and helical $p$-wave triplet components.
The pair potential of $d+p$ II is the order parameter discussed in the interfacial 
superconductivity~\cite{yada}. Although the pair potential of $d+p$ I may not have a 
relation to any materials, we consider it for theoretical interest.
The energy eigen values of Eq.~(\ref{bdg}) are $E=\pm E_{\pm}$ with 
$E_\pm=\sqrt{ (\xi_{\boldsymbol{k}} \pm \lambda)^2 + \Delta_\pm^2}$ and
\begin{align}
\Delta_\pm(\gamma)=\left\{\begin{array}{ll}
 \Delta_{\text{S}} \pm \Delta_{\text{T}}, & s+p, \\
 \Delta_{\text{S}}\sin 2\gamma \pm \Delta_{\text{T}}, & d+p\; \text{I}, \\
 (\Delta_{\text{S}} \pm \Delta_{\text{T}})\sin 2\gamma, & d+p\; \text{ II}. 
\end{array}\right.
\end{align}
To represent the wave function of a quasiparticle, we need  
another values of the pair potential defined by 
$\tilde{\Delta}_\pm(\gamma)={\Delta}_\pm(\pi-\gamma)$
\begin{align}
\tilde{\Delta}_\pm(\gamma)=\left\{\begin{array}{ll}
 \Delta_{\text{S}} \pm \Delta_{\text{T}}, & s+p, \\
 -\Delta_{\text{S}}\sin 2\gamma \pm \Delta_{\text{T}}, & d+p\; \text{I}, \\
 -(\Delta_{\text{S}} \pm \Delta_{\text{T}})\sin 2\gamma, & d+p\; \text{ II}. 
\end{array}\right.
\end{align}

\begin{widetext} 
The wave function in the left superconductor 
is obtained as
\begin{align}
&\Psi_L(x,y) =  \check{\Phi}_L
\left[ 
\left[
\begin{array}{cc}
u_+ & u_- \\
-ie^{i\gamma} u_+ & ie^{i\gamma}u_- \\ 
ie^{i\gamma} v_ + & -ie^{i\gamma} v_-\\ 
v_+ & v_- 
\end{array} 
\right]
\left[ \begin{array}{c}a_+ \\ a_-\end{array} \right] e^{ik_xx}
+
\left[\begin{array}{cc} 
\tilde{v}_+ & \tilde{v}_- \\ 
ie^{-i\gamma}\tilde{v}_+ & -ie^{-i\gamma}\tilde{v}_- \\ 
-ie^{-i\gamma} \tilde{u}_+ & ie^{-i\gamma} \tilde{u}_-\\ 
\tilde{u}_+ & \tilde{u}_- \end{array} \right]
\left[ \begin{array}{c}b_+ \\ b_-\end{array}\right] e^{-ik_xx}
\right.\nonumber\\
&\left.+
\left[\begin{array}{cc} \tilde{u}_+ & \tilde{u}_- \\ 
ie^{-i\gamma}\tilde{u}_+ & -ie^{-i\gamma}\tilde{u}_- \\ 
-ie^{-i\gamma} \tilde{v}_+ & ie^{-i\gamma} \tilde{v}_-\\ 
\tilde{v}_+ & \tilde{v}_- \end{array} \right]
\left[ \begin{array}{c}A_+ \\ A_-\end{array}\right]e^{-ik_xx}
+
\left[\begin{array}{cc} {v}_+ & {v}_- \\ 
-ie^{i\gamma}{v}_+ & ie^{i\gamma}{v}_- \\ 
ie^{i\gamma} {u}_+ & -ie^{i\gamma} {u}_-\\ 
{u}_+ & {u}_- \end{array} \right]
\left[ \begin{array}{c}B_+ \\ B_-\end{array}\right]e^{ik_xx}
\right]e^{ik_yy}
,\\
&u_\pm =\sqrt{ \frac{1}{2}\left(1+\frac{\Omega_\pm}{E}\right)}, \;
v_\pm =\sqrt{\frac{1}{2}\left(1-\frac{\Omega_\pm}{E}\right)}s_\pm, \;
\tilde{u}_\pm =\sqrt{\frac{1}{2}\left(1+\frac{\tilde{\Omega}_\pm}{E}\right)},\;
\tilde{v}_\pm =\sqrt{\frac{1}{2}\left(1-\frac{\tilde{\Omega}_\pm}{E}\right)}
\tilde{s}_\pm, \\
&\Omega_\pm= \sqrt{ E^2- \Delta_\pm^2 }, \;
\tilde{\Omega}_\pm=\sqrt{E^2- \tilde{\Delta}_\pm^2}, \;
s_\pm= \frac{ \Delta_\pm}{|\Delta_\pm|}, \;
\tilde{s}_\pm= \frac{ \tilde{\Delta}_\pm}{|\tilde{\Delta}_\pm|}, 
\; \check{\Phi}_{j}=\text{diag}\left\{ e^{i\varphi_j/2},e^{i\varphi_j/2},
e^{-i\varphi_j/2},e^{-i\varphi_j/2} \right\},
\end{align}
where $a_\pm$ and $b_\pm$ are the amplitudes of incoming waves, 
$A_\pm$ and $B_\pm$ are the amplitudes of outgoing waves, and
 $\varphi_j$ for $j=L$ or $R$ is the macroscopic phase of a superconductor.  
In the same way, the wave function in the right superconductor 
is represented by 
\begin{align}
\Psi_R(x,y) = & \check{\Phi}_R
\left[ 
\left[
\begin{array}{cc}
u_+ & u_- \\
-ie^{i\gamma} u_+ & ie^{i\gamma}u_- \\ 
ie^{i\gamma} v_ + & -ie^{i\gamma} v_-\\ 
v_+ & v_- 
\end{array} 
\right]
\left[ \begin{array}{c}C_+ \\ C_-\end{array} \right] e^{ik_xx}
+
\left[\begin{array}{cc} 
\tilde{v}_+ & \tilde{v}_- \\ 
ie^{-i\gamma}\tilde{v}_+ & -ie^{-i\gamma}\tilde{v}_- \\ 
-ie^{-i\gamma} \tilde{u}_+ & ie^{-i\gamma} \tilde{u}_-\\ 
\tilde{u}_+ & \tilde{u}_- \end{array} \right]
\left[ \begin{array}{c}D_+ \\ D_-\end{array}\right] e^{-ik_xx}
\right]e^{ik_yy},\label{psi_right}
\end{align}
with $C_\pm$ and $D_\pm$ being amplitudes of outgoing waves.
At the junction interface, we introduce the potential barrier 
described by $V_0\delta(x)$. Throughout this paper, we fix 
$z_0\equiv (V_0m)/(\hbar^2 k_F)=5$, which leads to the transmission 
probability of the insulating barrier $T_B=\int_0^{\pi/2}d\gamma \cos^3\gamma/(z_0^2
+\cos^2\gamma)$ being about 0.01. 
By eliminating $C_\pm$ and $D_\pm$ using a boundary 
condition, it is possible to obtain the reflection coefficients, 
\begin{align}
\left[ \begin{array}{c} A_+ \\ A_- \\ B_+ \\ B_- \end{array} \right]
=\left[ \begin{array}{cc} \hat{r}_{ee} & \hat{r}_{eh} \\
\hat{r}_{he} & \hat{r}_{hh} \end{array}\right]
\left[ \begin{array}{c} a_+ \\ a_- \\ b_+ \\ b_- \end{array} \right].
\end{align}
The Josephson current can be calculated based on a formula~\cite{ya01} 
after applying the continuation $E\to i\omega_n=i(2n+1)\pi T$,
\begin{align}
J=\frac{e}{2\hbar}\sum_{k_y}T\sum_{\omega_n}
\textrm{Tr}\left[
\left[ \begin{array}{cc}\frac{\Delta_+}{\Omega_{n+}} & 0 \\
0 & \frac{\Delta_-}{\Omega_{n-}} \end{array} \right] \hat{r}_{he}
-
\left[ \begin{array}{cc}\frac{\tilde{\Delta}_+}{\tilde{\Omega}_{n+}} & 0 \\
0 & \frac{\tilde{\Delta}_-}{\tilde{\Omega}_{n-}} \end{array} \right] \hat{r}_{eh}
\right],
\end{align}
with $\Omega_{n\pm}=\sqrt{ \omega_n^2+\Delta_\pm^2}$ and 
$\tilde{\Omega}_{n\pm}=\sqrt{ \omega_n^2+\tilde{\Delta}_\pm^2}$.
We introduce a parameter $0\leq \alpha\leq 1$ 
to tune the mixing rate between the spin-singlet and spin-triplet components 
as 
\begin{align}
\Delta_{\text{S}}= \alpha\Delta, \;\;
\Delta_{\text{T}}=(1-\alpha)\Delta,\label{alphadef}
\end{align}
where the dependence of $\Delta$ on temperature $T$ is described by the BCS theory.
The energy spectra of subgap state at a surface of NCS is calculated from 
Eq.~(\ref{psi_right}) with a boundary condition $\Psi_R(0,y)=0$.
\end{widetext}

\section{Results}
\begin{figure}[ht]
\begin{center}
\includegraphics[width=9cm]{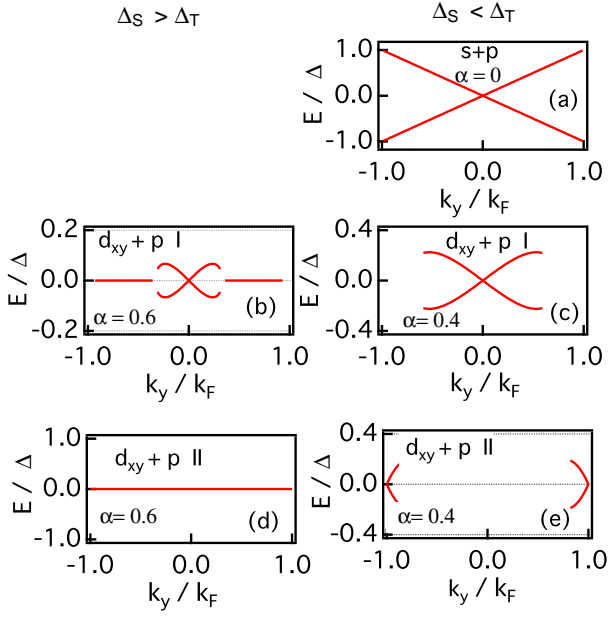}
\end{center}
\caption{(color online). The energy spectra of surface bound state.
(a): $s+p$ at $\alpha=0$, 
(b): $d+p $ I at $\alpha=0.6$, 
(c): $d+p $ I at $\alpha=0.4$,
(d): $d+p$ II at $\alpha=0.6$, and 
(e): $d+p$ II at $\alpha=0.4$.
The horisontal axis $k_y/k_F$ corresponds to $\sin\gamma$.
}\label{fig2}
\end{figure}

At first, we summarize the energy spectra of subgap state at a surface 
of superconductor in Fig.\ref{fig2}.
In $s+p$ mixture, energy of bound state satisfies
\begin{align}
(E^2-\Delta_+\Delta_-)\cos^2\gamma + \Omega_+\Omega_-(1+\sin^2\gamma)=0.
\end{align}
It has been already known that 
the surface bound state is absent for 
$\Delta_{\text{S}}>\Delta_{\text{T}}$, whereas the surface bound 
states with the linear dispersion exist for 
$\Delta_{\text{S}}>\Delta_{\text{T}}$
as shown in (a).

In $d+p$ I mixture, energy of the surface bound state satisfies
\begin{align}
E^2(1+\cos^2\gamma) + \sin^2\gamma(\Delta_+\Delta_- + \Omega_+\Omega_-)=0.
\end{align}
The equation has two solutions. The first one is $E=0$ which 
is allowed for $|\sin\gamma| \geq \Delta_{\text{T}}/\Delta_{\text{S}}$
as shown in Fig.~\ref{fig2}(b).
The dispersionless zero-energy bound states are a direct consequence 
of the $d_{xy}$-wave symmetry~\cite{tanaka95}.
Therefore such flat zero-energy state is absent for 
$\Delta_{\text{S}}<\Delta_{\text{T}}$.
The second solution is given by
\begin{align}
E=\pm\sin\gamma\sqrt{\Delta^2_{\text{T}}-\Delta^2_{\text{S}}4\sin^2\gamma}
\end{align}
for $|\tan\gamma|<\Delta_{\text{T}}/(2\Delta_{\text{S}})$ as shown in 
Fig.~\ref{fig2}(b) and (c).

 In $d+p$ II mixture, energy of the surface bound states 
 satisfies
 \begin{align}
E^2(1+\sin^2\gamma) + \cos^2\gamma(\Delta_+\Delta_- + \Omega_+\Omega_-)=0.
\end{align}
The equation has two solutions. The first one is $E=0$ 
for all $\gamma$ which 
is possible only when $\Delta_{\text{S}}>\Delta_{\text{T}}$
as shown in Fig.~\ref{fig2}(d).
The second solution is given by
\begin{align}
E=\pm 2 \cos^2\gamma \sqrt{\Delta^2_{\text{T}}\sin^2\gamma-\Delta^2_{\text{S}}},
\label{edp2}
\end{align}
which is allowed for 
$|\sin\gamma|<\sqrt{ \Delta_{\text{S}}/\Delta_{\text{T}}}$ as shown in 
Fig.~\ref{fig2}(e).

\subsection{s+p} 
\begin{figure}[tbh]
\begin{center}
\includegraphics[width=8cm]{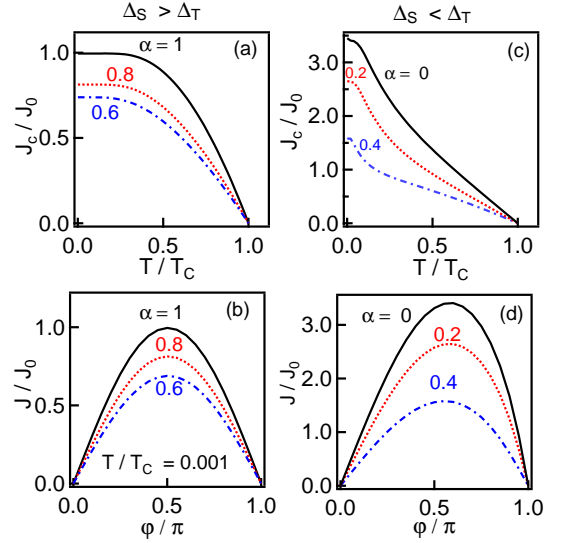}
\end{center}
\caption{(color online). Results for $s+p$ mixture.
The critical Josephson current is plotted as a function of temperature 
in (a) and (c). The current-phase relationships are show in (b) and (d) 
at a low temperature ($T=0.001T_c$). The transmission probability of the tunnel junction
$T_B$ is fixed about 0.01 throughout this paper.
 }
\label{fig3}
\end{figure}

In Fig.~\ref{fig3}, we show the calculated results of Josephson current for $s+p$ mixture.
In Fig.~\ref{fig3}(a), we plot the critical Josephson current 
as a function of
temperature for several choices of $\alpha$ satisfying
$\Delta_{\text{S}} > \Delta_{\text{T}}$. 
The Josephson current is normalized by $J_0=\pi \Delta_0/( 2eR_N)$, where 
$R_N$ is the normal resistance of the junction. In the case of metallic 
superconductor junctions, the Josephson critical current becomes $J_0$ at the 
zero temperature.
In Fig~\ref{fig3}(b), we show the current-phase relationship (CPR)
at a low temperature $T=0.001T_c$ for $\alpha=1$, 0.8 and 0.6, 
where $\varphi=\varphi_L-\varphi_R$ is the phase difference across the junction. 
The critical Josephson current saturates at low temperature and 
the CPR is sinusoidal for $\alpha=1$, 0.8 and 0.6.
Thus the Josephson current for $\Delta_{\text{S}} > \Delta_{\text{T}}$ 
obeys the Ambegaokar-Baratoff relation because there is no surface zero-energy 
states.
In Fig.~\ref{fig3}(c) and (d), we respectively show the dependence of critical current 
on temperature and the CPR at a low temperature 
for several choices of $\alpha$ satisfying 
 $\Delta_{\text{S}} < \Delta_{\text{T}}$. 
The results in Fig.~\ref{fig3}(c) and (d) show qualitatively different behavior 
from those in Fig.~\ref{fig3}(a) and (b), respectively.
The critical Josephson current in Fig.~\ref{fig3}(c) increases with decreasing 
temperature even far below $T_c$. 
This behavior is called low-temperature anomaly 
of Josephson current. The resonant tunneling through the surface bound state 
at the zero energy is responsible for the anomaly~\cite{tanaka96,barash}.
Such zero-energy state is possible $\sin\gamma=0$ as shown in Fig.~\ref{fig2}(a).
According to previous papers~\cite{barash2,ya02}, 
the Josephson critical current increases 
logarithmically with decreasing temperature for $\alpha=0$. 
The results for $\alpha=0.2$ and 0.4 also show the logarithmic low-temperature 
anomaly. Correspondingly, the contribution of higher harmonics
slightly deviates the CPR from the sinusoidal relation as shown in Fig.~\ref{fig3}(d).
Thus the characteristic feature of Josephson current qualitatively changes 
at the singular point of $\Delta_{\text{S}} = \Delta_{\text{T}}$.

\subsection{d+p I} 
\begin{figure}[tbh]
\begin{center}
\includegraphics[width=8cm]{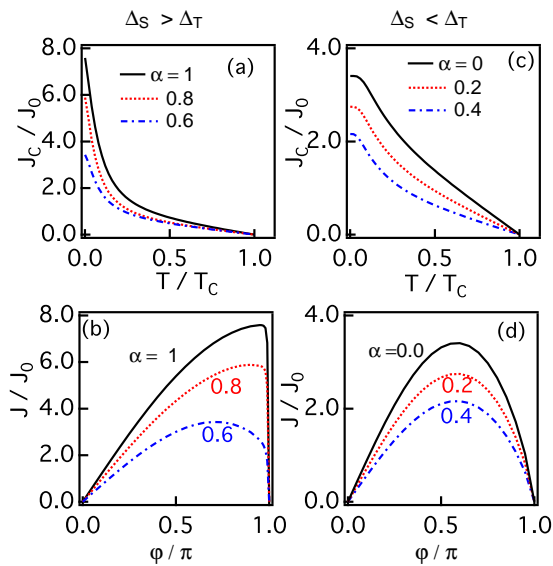}
\end{center}
\caption{(color online). Result for $d+p$ I.
The critical Josephson current is plotted as a function of temperature 
in (a) and (c). The current-phase relationship is show in (b) and (d) 
at low temperature 
$T=0.001T_c$.}
\label{fig4}
\end{figure}

Next, we show the calculated results of Josephson current for $d+p$ I mixture
as shown in Fig.~\ref{fig4}.
\begin{figure}[tbh]
\begin{center}
\includegraphics[width=8cm]{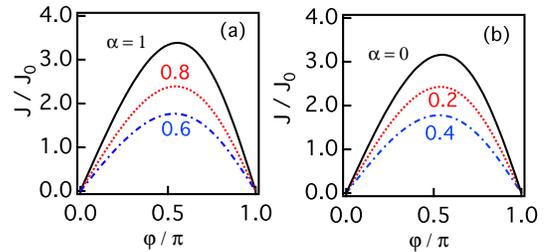}
\end{center}
\caption{(color online). The current-phase relationship in Fig. 4(b) and (d) 
are calculated for a higher temperature at $T=0.1T_c$.
}
\label{fig5}
\end{figure}
The temperature dependence of critical current for $\alpha=$1, 0.8, and 0.6 
satisfying
$\Delta_{\text{S}} > \Delta_{\text{T}}$
show the low-temperature anomaly as shown in Fig.~\ref{fig4}(a). 
The critical current increase as $T^{-1}$ with decreasing temperature
 at $\alpha=1$~\cite{tanaka96}.
The results for $\alpha=0.8$ and 0.6 also show such power-law like
low-temperature anomaly. 
Corresponding CPR shown in Fig.~\ref{fig4}(b) indicates 
jump at $\varphi=\pi$ because of the contributions of higher harmonics.
In this case, the surface bound state are energetically localized at
$E=0$ as shown in Fig.~\ref{fig2}(b). 
The resonant tunneling through such zero-energy states
causes the strong low-temperature anomaly.
For $\Delta_{\text{S}} < \Delta_{\text{T}}$, on the other hand, the Josephson current 
in Fig.~\ref{fig4}(c) and (d) have properties similar to those in 
Fig.~\ref{fig3}(c) and (d), respectively. 
The critical current indicates the logarithmic low-temperature anomaly.
The presence of the zero-energy surface bound state at 
$\sin\gamma=0$ in Fig.~\ref{fig2}(c) explains the similarity.
Thus the characteristic feature of Josephson current for $d+p$ I mixture
also qualitatively changes around the point of 
$\Delta_{\text{S}} = \Delta_{\text{T}}$.
 
 The large deviation of CPR from the sinusoidal function in Fig.~\ref{fig4}(b)
can be seen only at low temperature. According to an analytical expression of 
Josephson current for $\alpha=1$, the higher harmonics contribute to the Josephson 
current when the temperature is much smaller than $\sqrt{T_B}\Delta_0$. 
Here $T_B$ is the transmission probability
of the tunnel junction and is about 0.01 in the present calculation. 
In Fig.~\ref{fig5}, we show CPR for a higher temperature at $T=0.1T_c$.
In both Figs.~\ref{fig4}(a) and (b), the CPR deviates from the sinusoidal
relation only slightly at $T=0.1T_c$. On the other hand, 
the amplitudes of the Josephson current remain sufficiently larger value than $J_0$.

\subsection{d+p II } 
\begin{figure}[tbh]
\begin{center}
\includegraphics[width=8cm]{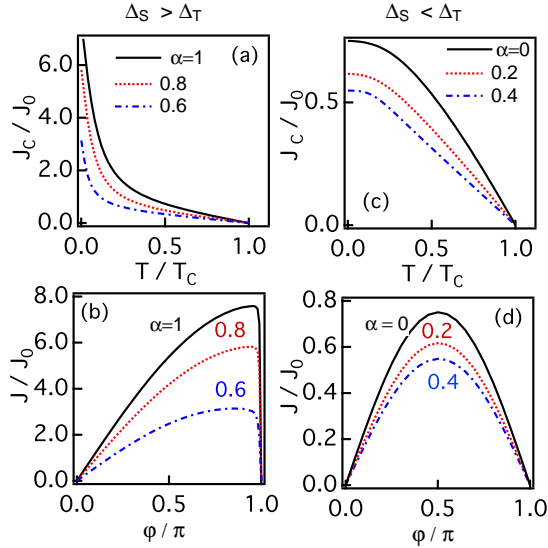}
\end{center}
\caption{(color online). Results for $d+p$ II.
The critical Josephson current is plotted as a function of temperature 
in (a) and (c). The current-phase relationship is show at low temperature 
$T=0.001T_c$.}
\label{fig6}
\end{figure}

Finally, we show the calculated results of Josephson current for $d+p$ II mixture
as shown in Fig.~\ref{fig6}.
The temperature dependence of critical current for several $\alpha$
satisfying $\Delta_{\text{S}} > \Delta_{\text{T}}$
indicate the strong low-temperature anomaly as shown in Fig.~\ref{fig6}(a) and (b). 
The critical current increase with decreasing temperature as $T^{-1}$ 
and the CPR at a low temperature shows the jump at $\varphi=\pi$.
The zero-energy surface bound states are possible for all $\gamma$ as shown 
in Fig.~\ref{fig2}(d).
The presence of the flat zero-energy states explains the 
similarity of the results in Fig.~\ref{fig6}(a) and (b) 
 to those shown in Fig.~\ref{fig4}(a) and (b), respectively.
The calculated results for $\alpha=$0.4, 0.2 and 0 satisfying 
 $\Delta_{\text{S}} < \Delta_{\text{T}}$ are shown in Fig.~\ref{fig6}(c) and (d).
On the contrary, the results for $\Delta_{\text{S}} < \Delta_{\text{T}}$
in Fig.~\ref{fig6}(c) and (d) 
has properties similar to those in Fig.~\ref{fig3}(a) and (b). 
Namely, the Josephson current saturates at low temperature and the CPR 
is sinusoidal at low temperature.
The zero-energy state at $\sin\gamma=\pm 1$ exists as 
shown in Fig.~\ref{fig2}(e).
 Although this zero-energy state appears as a results of the resonant 
 Andreev reflection~\cite{ya04}, 
it does not so much affect the Josephson current. 
The wavenumber $\sin\gamma=\pm 1$ means 
$k_x=\cos\gamma=0$. Thus a quasiparticle does not have momenta 
in the current direction in the zero-energy state. 
When we consider huge spin-orbit coupling, 
it has been pointed out that~\cite{tanaka10,yada11,sato11} the flat zero-energy states appear
for $\sin\gamma > (1- 2 \lambda/\mu)$. 
In such case, the flat zero-energy state may cause the low-temperature 
anomaly. 
This statement, however, is still unclear 
in realistic junctions with a thick insulating barrier because 
the contribution of a quasiparticle with $\sin\gamma \approx \pm 1$ 
to Josephson current becomes exponentially small. 
Within the approximation of $\lambda/\mu\ll 1$, there is no 
effective zero-energy state which causes the low-temperature 
anomaly for $\Delta_{\text{S}} < \Delta_{\text{T}}$. 
Therefore the Josephson current in Fig.~\ref{fig6}(c) and (d) show 
qualitatively the same behavior 
as those in Fig.~\ref{fig3}(a) and (b), respectively.
Thus the characteristic feature of Josephson current for $d+p$ II 
also qualitatively changes around the point of 
$\Delta_{\text{S}} = \Delta_{\text{T}}$.

At $\Delta_{\text{S}}=0$, subgap state with the linear dispersion 
appears around $\sin\gamma=0$ as mathematically shown in Eq.~(\ref{edp2}). 
This zero-energy state, however, is not a result of the resonant 
Andreev reflection~\cite{ya04} but is a result of a node in 
the pair potential. Thus $\Delta_{\text{S}}=0$ cannot be a critical 
point. 
Actually, the results for $\alpha=0$ in Fig.\ref{fig6}(c) 
 show the saturation of the Josephson critical current at low 
 temperature.

\section{conclusion}
In summary, we have theoretically studied the 
Josephson current between two noncentrosymmetric superconductors 
based on the Bogoliubov-de Gennes equation and a general current formula.
We have assumed three types of order parameter which consists 
of spin-singlet $\Delta_{\text{S}}$ and spin-triplet components $\Delta_{\text{T}}$ 
at the same time. The Josephson current for $\Delta_{\text{S}}>\Delta_{\text{T}}$
shows clearly the different 
characteristic behavior from those for $\Delta_{\text{S}}<\Delta_{\text{T}}$
for all pairing symmetries. The clear difference can be understood by analyzing
the topologically protected zero-energy states at a surface of noncentrosymmetric 
superconductor.
The dispersionless zero-energy bound states are responsible for 
strong low-temperature anomaly of Josephson current in which 
the Josephson critical current increases as $1/T$ with decreasing temperature.
The surface state with linear dispersion causes the weak low-temperature 
anomaly in which the Josephson critical current increases 
logarithmically with decreasing temperature. 
When the surface zero-energy state is absent, the Josephson current 
obeys the Ambegaokar-Baratoff formula.

\section{acknowledgement}
This work was supported by KAKENHI(No. 22540355) and 
the "Topological Quantum Phenomena" (No. 22103002) Grant-in Aid for Scientific Research on Innovative Areas from the Ministry of Education, 
Culture, Sports, Science and Technology (MEXT) of Japan.

\end{document}